\documentclass{appolb}
\usepackage{epsfig}

\def\bge{\begin{equation}}
\def\ene{\end{equation}}
\def\bgea{\begin{eqnarray}}
\def\enea{\end{eqnarray}}
\def\nn{\nonumber}

\def\lo{$\mathrm{LO}$}
\def\nlo{$\mathrm{NLO}$}
\def\chpt{$\mathrm{\chi PT}$}
\def\rxt{$\mathrm{R\chi T}$}

\def\dd{\mathrm{d}}
\begin{document}
         \date{October 26, 2009}                                                                         %
        \pagestyle{plain}                                                                    %

\title{ RADIATIVE RETURN: A PROGRESS ON FSR TESTS%
	\thanks{Presented by S.~Ivashyn at Topical workshop: 
	{\sl Low energy constraints
	on extensions of the Standard Model}, 23-27 July 2009,
	Kazimierz, Poland 
	and 	
	XXXIII International Conference of Theoretical Physics
	{\sl MATTER TO THE DEEPEST: Recent Developments in Physics
	of Fundamental Interactions}, 11-16 September 2009,
	Ustro\'n, Poland
	}%
}
\author{
	Sergiy~IVASHYN
	\address{Institute of Physics, University of Silesia, Katowice PL-40007, Poland
	\\and
	NSC ``KIPT'', Kharkov UA-61108, Ukraine}
\\\vskip8pt 
	Henryk~CZY\.Z
	\address{Institute of Physics, University of Silesia, Katowice PL-40007, Poland}
\\\vskip8pt 
	Alexandr~KORCHIN
	\address{NSC ``KIPT'', Kharkov UA-61108, Ukraine}
}

\maketitle
\begin{abstract}
	To improve the accuracy of the
	$e^+e^- \to \pi^+\pi^-\gamma$ Radiative Return method,
	one has to control the theoretical uncertainty of
	the final-state photon emission. 
	It is of particular importance at DAPHNE for the analysis,
	where cuts are relaxed to cover the threshold region.
	By means of Monte Carlo generator PHOKHARA we compare several
	final-state radiation models and present
	results, relevant for a meson factory 
	running at $\sqrt{s}=1$~GeV. 
\end{abstract}


\section{Introduction}
\label{intro}

The Radiative Return Method~\cite{Chen:1974wv,Arbuzov:1998te,Binner:1999bt}  (RRM)
allows an extraction of the hadronic cross section $\sigma^{\mathnormal{had}} (Q^2)$ 
for hadronic invariant mass squared $Q^2$ from the energy threshold up to
the nominal energy of the experiment at
the fixed beam energy $e^+ e^-$ colliders.
High-luminosity meson 
factories are especially suited for this purpose~\cite{Grzelinska:2008eb}.
Interest in {\it precise} measurement of $\sigma^{\mathnormal{had}} (Q^2)$
is motivated, in part, by its relevance to the hadronic contribution 
to the muon anomalous magnetic moment~$a_\mu^{had}$~\cite{Jegerlehner:2009ry,Prades:2009qp}
and the electromagnetic fine structure constant~\cite{Jegerlehner:2008rs}. 
However, the method can also be applied
to extract the meson form factors
and other meson properties.

The RRM uses
$\dd\sigma(e^+e^-\to \mathnormal{hadrons} +\mathnormal{photons})/ \dd Q^2$,
 a measured differential cross section,
for the extraction of $\dd \sigma(e^+e^-\to \mathnormal{hadrons})$. 
For the theoretical description, the perturbative QED diagrams 
at the leading order in QED coupling $\alpha$ (\lo) and
 at the next to leading order (\nlo) are considered and
 classified as
 initial-state radiation (ISR) or final-state radiation (FSR) ones.
 The kinematic cuts are applied to
 sufficiently  suppress the FSR, whenever possible, as the 
 factorization
 $\left. \dd\sigma(s,Q^2) \right|_{ISR} = \mathcal{R}(s,Q^2) \;\times\;\dd\sigma_{\mathnormal{had}} (Q^2)$,
 which allows for $\dd \sigma^{\mathnormal{had}} (Q^2)$ extraction,
 holds for diagrams with ISR photons only.
 The function $\mathcal{R}(s,Q^2)$ is given by QED.
The FSR part is model-dependent,
thus dedicated numerical studies are needed for correct 
ISR-FSR separation.
The Monte Carlo generator PHOKHARA was
developed for these and related purposes:
FSR at \nlo\ has been included~\cite{Czyz:2003ue} for pion pair production and,
in addition to scalar QED (sQED), some particular ingredients (the $\phi$ radiative decay) 
were implemented~\cite{PHOKHARA:phi}.
The FSR was also examined by 
other Monte Carlo programs, e.g., that
with the Resonance Chiral Theory (\rxt) motivated framework~\cite{Pancheri:2006cp}
and phenomenologically-oriented model~\cite{Shekhovtsova:2009yn},
which was also included into PHOKHARA~6.1~\cite{phokhara:web}.

The reaction $e^+ e^- \to \pi^+ \pi^- \gamma$ was explored  
by KLOE~\cite{kloe}:
the cross section $\dd \sigma^{\mathnormal{had}}/\dd Q^2$ 
and pion form factor $F_\pi(Q^2)$ 
in the range $0.35$~GeV$^2 < Q^2 < 0.95$~GeV$^2$ were extracted~\cite{KLOE:RR:fpi}
from the on-peak ($\sqrt{s}=M_\phi=1.02$~GeV) data
sample by means of the RRM.
However, the kinematic cuts, which were applied in order
to suppress FSR, did not allow to measure at $Q^2$ below $0.35$~GeV$^2$.

One can measure the $F_\pi(Q^2)$ in the threshold region relaxing some of the cuts,
but then one has to {\it subtract the FSR contribution}.
In this scenario, one needs to control the 
description of the final-state emission process
and detailed studies are needed to estimate the theoretical uncertainty.
To simplify the analysis, it is better to perform the measurement 
off the $\phi$ meson peak, because in this case the contributions
from the $\phi$ meson radiative decays are small and the FSR models 
can be controlled easier.

The investigations presented here are of particular importance
for the forthcoming KLOE RRM analysis of pion pair production. 
We focus on the off-$\phi$-peak measurement, at $e^+ e^-$ center-of-mass energy $\sqrt{s}=1$~GeV,
for which KLOE collected $230$~pb$^{-1}$ of data~\cite{beltrame:phd}.
%
Due to the interest in precision
at small $Q^2$ (i.e., below the $\rho$ resonance), 
the Chiral Perturbation Theory (\chpt)~\cite{Gasser:1983yg}
can be helpful. The relevant theoretical aspects are sketched in Section~\ref{section_theory}.
We use Monte Carlo generator PHOKHARA to compare several
final-sate radiation models.
The theoretical heritage of Virtual Compton Scattering (VCS) off the pion
in \chpt\ framework~\cite{Unkmeir:1999md,Fuchs:2000pn} is used
to estimate the r\^ole of higher order \chpt\ effects.

The numerical results for cross section and asymmetry
are presented in Section~\ref{section_numer}.
All the parameters of implemented models
are fixed independently, thus one deals with model {\it predictions}.
In Section~\ref{section_conclus} we present our conclusions.


%
\section{Theoretical issues of final-state radiation}
\label{section_theory}
The transition $\gamma^\ast \to \pi^+\pi^-\gamma$
is described by the model-dependent FSR tensor $M^{\mu\nu}$\footnote{
For example, the matrix element of $e^+e^- \to \gamma^\ast \to \pi^+\pi^-\gamma$ 
reads:\\ $M^{(LO)}_{FSR}= s^{-1} e\; \bar{v}\;\gamma_\mu u \;\epsilon^\ast_{\nu} \; M^{\mu\nu}$,
where $e = \sqrt{4\pi \alpha}$.}.
In all realistic models it contains the same Born-level contribution 
$M_{Born}^{\mu\nu}$, which corresponds to a no-structure approximation for pion (Scalar QED or lowest-order \chpt).
Thus we consider $M_{Born}^{\mu\nu}$ as a model-independent part.


The first correction accounts for the pion
structure by means of the pion form factor. 
It replaces the Born-level amplitude by ``Generalized Born'' (GB) 
one~\cite{Unkmeir:1999md}, $M_{GB}^{\mu\nu}$,  which is also 
called ``sQED*VMD''~\cite{Czyz:2003ue}.
Generalized Born FSR tensor reads
\begin{equation}
\label{eq:GB}
M_{GB}^{\mu\nu}\!= - \mathrm{i}e^2 F_\pi(P^2)
\left( \frac{(k+q_1-q_2)^\mu q_1^\nu}{q_1\cdot k} 
+       \frac{(k+q_2-q_1)^\mu q_2^\nu}{q_2\cdot k}
- 2\ g^{\mu\nu}\!
\right),
\end{equation}
where $P$ and $k$ are the virtual and real photon momenta,
$q_1$ and $q_2$ --- pion momenta ($Q = q_1+q_2$).

Limit $F_\pi(P^2) \to 1$ reproduces the
$M_{Born}^{\mu\nu}$ amplitude.
Notice, that in $e^+e^- \to \pi^+\pi^-\gamma$ at \lo, 
$P^2 = s$, thus the form factor $F_\pi(P^2)\ne 1$
and its correction is never negligible.
Also this part is well established both theoretically and experimentally.

In the ISR amplitude, with $\gamma^\ast \to \pi^+\pi^-$ transition
in the final state, one finds $F_\pi(Q^2)$ factor in the amplitude.
Therefore, the $\pi^+\pi^-$ invariant mass distribution is 
governed by the form factor shape.
For consistency, one has to 
use the same expression for the pion form factor
in the ISR and FSR amplitudes. 
It is important to take the form factor tested experimentally
and not to rely only on a particular model assumptions.
This will be illustrated in the next Section.
In order to understand the accuracy of $M_{GB}^{\mu\nu}$ approximation,
we study further corrections
using the models of Refs.~\cite{Shekhovtsova:2009yn,Unkmeir:1999md,Fuchs:2000pn}.


The first model, ``VMD*\chpt'',
is based on $\mathcal{O}(p^4)$ \chpt\
$SU(2)$ description of VCS 
$\gamma^\ast \pi^\pm  \to \gamma \pi^\pm $~\cite{Unkmeir:1999md}
and that in $SU(3)$ case~\cite{Fuchs:2000pn}.
The FSR tensor has the form $M^{\mu\nu} = M_{GB}^{\mu\nu} + M_{NB}^{\mu\nu}$.
The first term is given by Eq.~(\ref{eq:GB})
and a straightforward improvement beyond \chpt\ 
is supposed (denoted by prefix ``VMD*''): 
the pion form factor $F_\pi$ is 
an external input (e.g., defined by parametrization
of the measured $F_\pi$).
The second term, {\it the Non-Born correction} to FSR reads:
$M_{NB}^{\mu \nu } = -ie^{2}\; (k^{\mu }Q^{\nu }-g^{\mu \nu }\;k\!\cdot\! Q) \; f^{NB}_{1}$,
where
\bgea  
f^{NB}_{1} &=& \frac{-1}{16\;\pi^2\; F^2}
\left( \frac{2}{3}(\bar{l}_6 - \bar{l}_5) 
+ \frac{P^2 - 2\;P\!\cdot\! k}{P\cdot k} 
\times \mathcal{G}_\pi \right),
\\
f^{NB}_{1} &=& \frac{-1}{16\;\pi^2\; F_0^2}
\!\left(\! 128 \pi^2(L^r_9 + L^r_{10}) + \frac{P^2 \! - \! 2\;P\!\cdot\! k}{P\cdot k}
\times\left( \mathcal{G}_\pi + \frac{1}{2}\mathcal{G}_K\right) \! \right)
\enea
in $SU(2)$ and $SU(3)$ framework, correspondingly; 
see original papers~\cite{Unkmeir:1999md,Fuchs:2000pn}
 for the explicit form of the loop functions
$\mathcal{G}_\pi$ and $\mathcal{G}_K$.
Numerical values of the low energy constants are
$F = 92.4$~MeV,
$(\bar{l}_6 - \bar{l}_5) = 3.0 \pm 0.3$ and
$F_0 = 87.7$~MeV, as cited in~\cite{Bijnens:2009pq},
and $(L^r_9 + L^r_{10}) = (1.32 \pm 0.14)\times10^{-3}$ at scale 
$\mu=M_\rho$, as estimated in~\cite{Unterdorfer:2008zz}.

 \begin{figure} \begin{center}
 \resizebox{1.0\textwidth}{!}{%
      \includegraphics{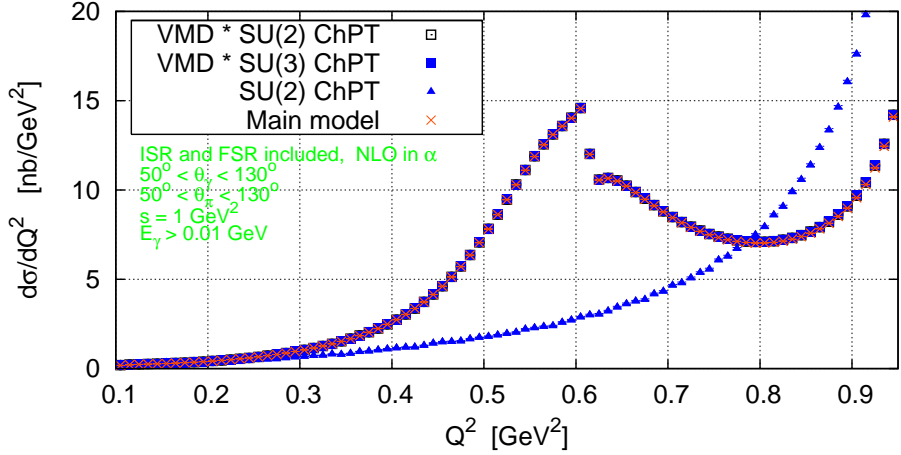} } 
      \end{center}
 \caption{ Differential cross section for $e^+e^- \to \pi^+ \pi^- \gamma$.
 All the listed models, but the one with the $\mathcal{O}(p^4)$
  \chpt\ form factor, give very close predictions (overlapping curves in the plot)
           }
 \label{fig:cs:cs}
 \end{figure}

The second model~\cite{Shekhovtsova:2009yn}, called the
``main model'' further in the text, can be considered as {\it a parametrization}  
of  $\pi^0\pi^0 \gamma$ KLOE data, transformed to $\pi^+\pi^- \gamma$
via isospin symmetry~\cite{Shekhovtsova:priv}.
It was implemented in 
FASTERD Monte Carlo generator~\cite{Shekhovtsova:2009yn}
and in PHOKHARA~6.1 recently~\cite{phokhara:web}.
The FSR tensor contains $M_{GB}^{\mu\nu}$ given by Eq.~(\ref{eq:GB})
and the Non-Born corrections due to important 
vector-resonance and double-vector-resonance contributions.

\begin{figure} \begin{center}
 \resizebox{1.0\textwidth}{!}{%
      \includegraphics{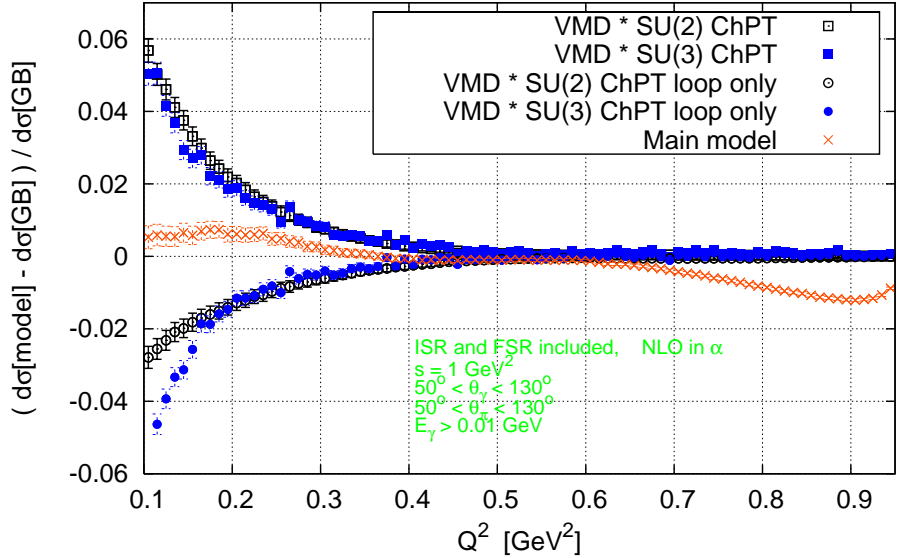} } 
 \end{center}
 \caption{ Non-Born corrections to $\dd \sigma/\dd Q^2$ (left)
           Note, that at \nlo\ the interference is neglected. At $Q^2 > 0.5$~GeV$^2$ 
	   all the curves but ``main model'' overlap}
 \label{fig:cs:nlonb}
 \end{figure}

%
\section{Numerical results}
\label{section_numer}

We use Monte Carlo generator PHOKHARA to compare 
the model-dependent effects in 
$e^+e^- \to \pi^+ \pi^- \gamma$ cross section and asymmetry
for the off-peak case at $\phi$-meson factory, $\sqrt{s}=1$~GeV.

First of all, we stress that 
any simplification of the pion form factor $F_\pi$
can drastically affect the model results.
Figure~\ref{fig:cs:cs} shows that
rigorous $\mathcal{O}(p^4)$ \chpt\ form factor~\cite{Gasser:1983yg}
gives completely wrong estimate for differential cross section
even in the region of $Q^2$ below the $\rho$ meson peak.
The theoretical explanation of the form factor r\^ole was given above.  
The form factor used in VMD*\chpt\  and ``main model''
is the parametrization of 
available data given by Gounaris-Sakurai version 
of Ref.~\cite{Bruch:2004py}.

\begin{figure} \begin{center}
 \resizebox{1.0\textwidth}{!}{%
      \includegraphics{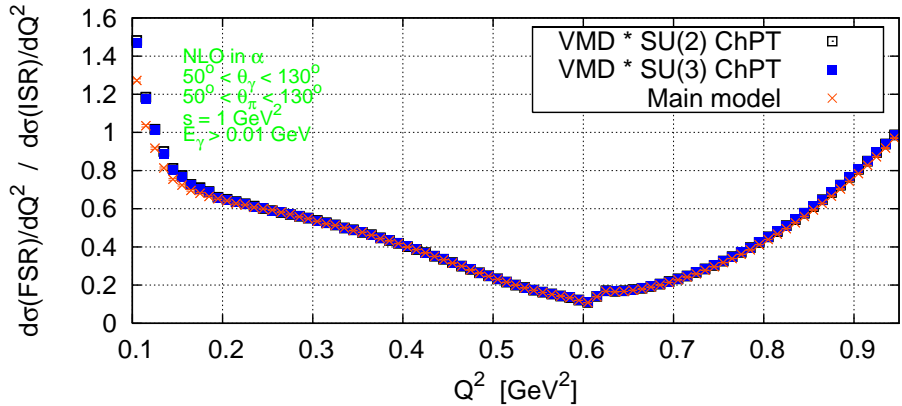} } 
 \end{center}
 \caption{ The r\^ole of FSR in $\dd \sigma/\dd Q^2$ (right). All the listed models 
          give very close predictions (overlapping curves in the plot).
          Note, that at \nlo\ the interference is neglected}
 \label{fig:cs:nlofsrisr}
 \end{figure}

In Fig.~\ref{fig:cs:cs}, one can see 
the very close cross section predictions,
despite the fact, that the models have {\it completely different}
Non-Born corrections.
This is due to the fact that the GB contribution dominate 
for the given event selection.
Taking the GB approximation, Eq.~(\ref{eq:GB}),
as a reference, we plot
$(\dd \sigma[model] - \dd \sigma[GB]) / \dd \sigma[GB]$.
To show the relative contribution of loop and ``constant''
terms in \chpt\ we consider also the case of
$(\bar{l}_6 - \bar{l}_5)$ and $(L^r_9 + L^r_{10})$ 
being artificially set to zero. 
Corresponding results are marked as ``loop only'' in the pictures.
Figure~\ref{fig:cs:nlonb} shows that 
the Non-Born corrections 
are at a few per cent level. 
From Fig.~\ref{fig:cs:nlofsrisr}
one concludes that the FSR contribution to the cross section is significant 
in the whole range of $Q^2$, especially at low $Q^2$.

 Pion forward-backward asymmetry (FBA) as
a function of $Q^2$ reads 
\bge
\nn A_{FB}(Q^2) = \frac{N(\theta_{\pi^+} > 90^o) - N(\theta_{\pi^+} < 90^o)} 
{N(\theta_{\pi^+} > 90^o) + N(\theta_{\pi^+} < 90^o)}(Q^2)
\ene
in terms of numbers of events.
Origin of the non-zero FBA is the interference of C-odd and C-even amplitudes,
e.g., that of ISR and FSR at \lo. 
Thus, FBA is sensitive to the relative phase, which 
may differ among the models even if they predict the same cross section. 
Notice, that the experimental data on asymmetry and cross section are 
to large extent independent.
Therefore the FBA is a good test for models.
Aspects of using the FBA in 
$e^+e^- \to \pi^+\pi^-\gamma$ were discussed
in~\cite{Binner:1999bt,Czyz:2003ue,PHOKHARA:phi,Pancheri:2006cp,our_2005}.

\begin{figure} \begin{center}
 \resizebox{1.0\textwidth}{!}{%
      \includegraphics{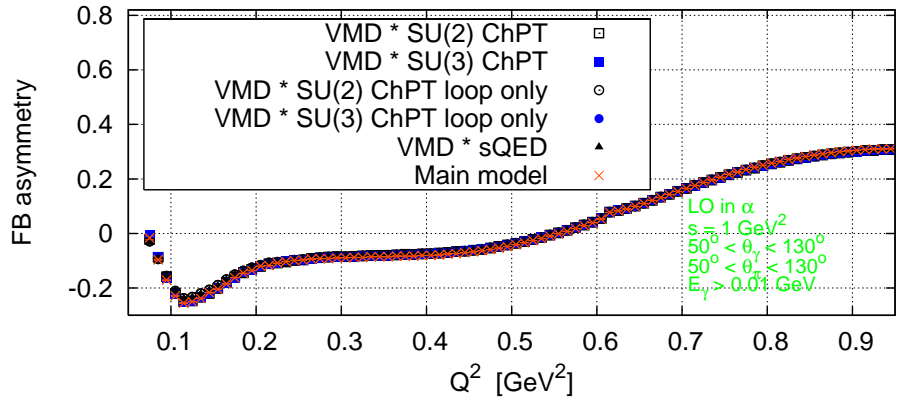} 
      } 
 \end{center}
 \caption{Forward-backward asymmetry for pion. All the listed models 
          give very close predictions (overlapping curves in the plot)}
 \label{fig:fba}
 \end{figure}
\begin{figure} \begin{center}
\resizebox{1.0\textwidth}{!}{%
      \includegraphics{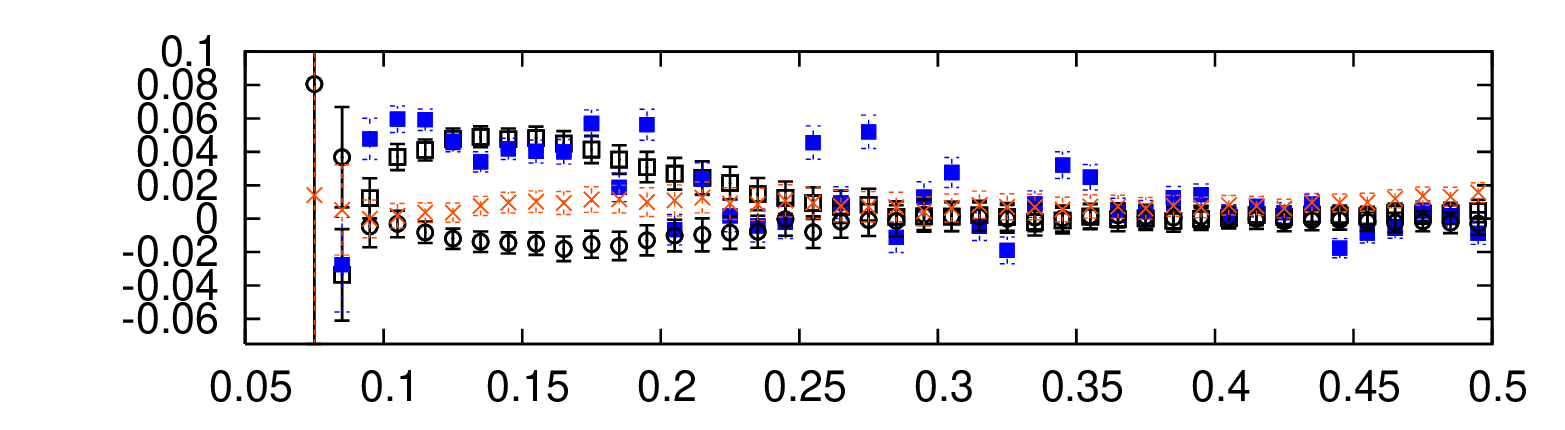} 
      } 
      
\resizebox{1.0\textwidth}{!}{%
      \includegraphics{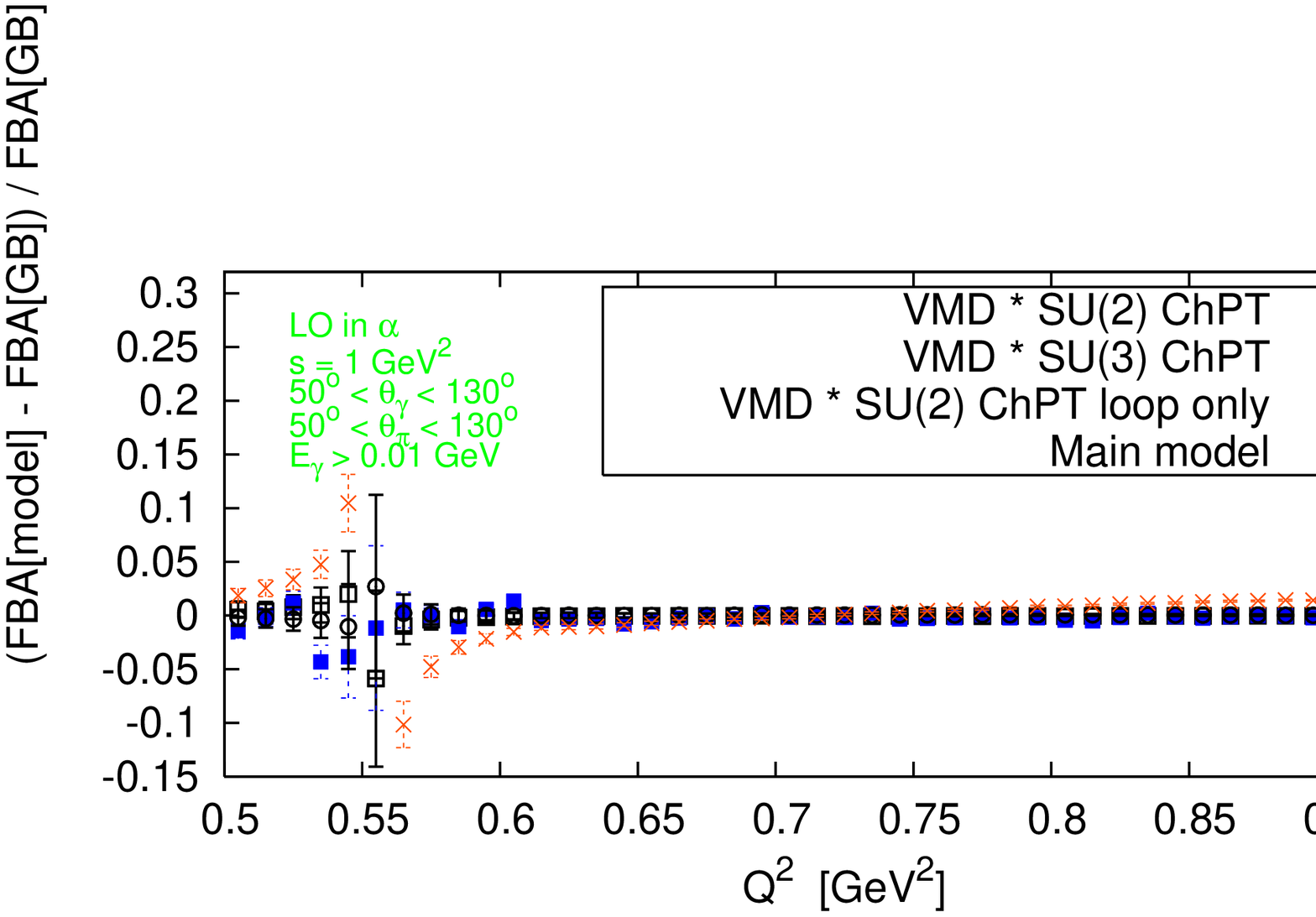} 
      } 
 \end{center}
 \caption{Relative non-born corrections to the forward-backward asymmetry for pion}
 \label{fig:fba:nb}
 \end{figure}

Figure~\ref{fig:fba} shows that FBA is sizable and relatively easy 
measurable.
From Fig.~\ref{fig:fba:nb} we conclude that the Non-Born corrections
to the FBA are of few per cent order and will not have a big influence on
the theoretical uncertainty.

It has to be stressed that if the \chpt\ corrections were not accounted for
in the formulae used to measure the $|F_\pi|$, they are partly
accounted for in the experimental parameters of $F_\pi$
and other model parameters. In other words,
one model should be used in all experimental analyzes and adding {\it ad hoc}
additional corrections is not appropriate.

%
\section{Conclusions}
\label{section_conclus}

Using PHOKHARA, we studied the corrections given by \chpt~\cite{Unkmeir:1999md,Fuchs:2000pn},
and by a phenomenological model including miscellaneous hadronic resonance effects~\cite{Shekhovtsova:2009yn}. 
Corrections due to $a_1$ resonance~\cite{our_2005} 
are to be considered elsewhere.
The r\^ole of the pion form factor is seen to be very important.
Final-state radiation is significant
in the whole range of $Q^2$, especially at low~$Q^2$.
We have found the \nlo\ corrections 
to be non-negligible,
even if the Generalized Born contribution is dominant.
Non-Born corrections are of order of few per cent.
They differ among the models, but 
it will be difficult to distinguish them with
the present KLOE off-peak statistics.
The results presented here show that one should include the 
\chpt\ corrections in the analysis when the accuracy of the
experiment reaches a per cent level.

{
\vspace{5pt}
\begin{center}{\bf Acknowledgments}\end{center}

We would like to thank Achim Denig, Stefan Scherer, Stefan M\"uller and Roman Zwicky
for discussions.
Partial support from MRTN-CT-2006-035482 ``FLAVIAnet'',
MRTN-CT-2006-035505 ``HEPTOOLS'' and Contract No.~227431 ``TARI''
is acknowledged.
}
\normalsize
%
%

%
%

\end{document}